\documentclass[preprint2]{aastex}
\usepackage[]{natbib}

\newcommand{\f}{4U\,1626--67 \,}
\newcommand{\fermigbm}{ {\it Fermi}/GBM }
\newcommand{\swiftbat}{{\it Swift}/BAT }

\newcommand{\be}{\begin{equation}}
\newcommand{\ee}{\end{equation}}
\newcommand{\bdm}{\begin{displaymath}}
\newcommand{\edm}{\end{displaymath}}

\shorttitle{New torque reversal and spin--up of 4U\,1626--67}
\shortauthors{Camero-Arranz et al.}

\begin{document}
\title{NEW TORQUE REVERSAL AND SPIN-UP OF 4U\,1626--67 OBSERVED BY FERMI/GBM AND SWIFT/BAT}

\author{A. Camero--Arranz\altaffilmark{1,2,3}, M.H. Finger\altaffilmark{1,4},  N.R. Ikhsanov\altaffilmark{5,6}, C.A. Wilson--Hodge\altaffilmark{1,5} and  E.Beklen\altaffilmark{7,8}}

\altaffiltext{1}{National Space Science and Technology Center, 320 Sparkman Drive, Huntsville, AL 35805}
\altaffiltext{2}{Fundaci\'{o}n Espa\~{n}ola de Ciencia y Tecnolog\'{i}a, C/Rosario Pino,14-16, 28020- Madrid, Spain}
\altaffiltext{3}{MICINN (Ministerio de Ciencia e Innovaci\'{o}n), C/Albacete, 5, 28027, Madrid, Espa\~{n}a}
\altaffiltext{4}{Universities Space Research Association, 6767 Old Madison Pike, Suite 450, Huntsville, AL 35806}
\altaffiltext{5}{NASA/Marshall Space Flight Center}
\altaffiltext{6}{Pulkovo Observatory, 196140, St. Petersburg, Russia}
\altaffiltext{7}{Physics Department, Middle East Technical University, 06531 Ankara, Turkey}
\altaffiltext{8}{Physics Department, S\"uleyman Demirel University, 32260 Isparta, Turkey}

\begin{abstract}
After about  18 years of steadily spinning down,  the accretion--powered pulsar 4U\,1626-67, experienced a new torque reversal at the beginning of 2008. For the present study we have used all available \fermigbm data since its launch in 2008 June 11 and over 5 yr of hard X-ray \swiftbat observations (starting from 2004 October up to the present time). From 2004 up to the end of 2007   the spin--down rate averaged at a mean rate of  $\sim \dot{\nu}=-4.8 \times10^{-13}$\,Hz\,s$^{-1}$ until the torque reversal reported here. This second detected torque reversal was centered near MJD 54500 (2008 Feb 4) and it lasted  approximately  150 days. During the reversal the source also underwent  an increase in flux by a fraction of $\sim$2.5.  Since then it has been  following a  steady spin--up at a mean rate of $\sim \dot{\nu}=4 \times 10^{-13}$\,Hz\,s$^{-1}$. We present a detailed long-term timing analysis of this source and a long term spectral hardness ratio study in order to see  whether there are spectral changes around this new observed torque reversal.
\end{abstract}


\section{INTRODUCTION}

 The accreting--powered pulsar \f was discovered
by {\it Uhuru} \citep{Giacconi72}. 
This low mass X--ray binary (LMXB)   consists of a 7.66 s X--ray pulsar accreting
from an extremely low mass companion (0.04 M$\odot$ for {\it i} = 18$^o$)
\citep{Levine88}. Although orbital motion has never been detected  in the  X--ray data, pulsed optical
emission reprocessed on the surface of the secondary revealed \cite[]{Middledich81}
the 42 min orbital period, confirmed by \cite{Chakrabarty98}.  Most likely the binary system  contains a hydrogen--depleted secondary to reach such a short orbital period
\citep{Paczynski&Sienkiewicz81}. The faint optical counterpart (KZ TrA, {\it V}$\sim$17.5)
has a strong UV excess and high optical pulse fraction \citep{McClintock77,McClintock80}.
A persistent 48 mHz quasi-periodic oscillation (QPO) has been detected in the X--ray emission
\citep{Shinoda90, KommersChakrabarty&Lewin98}. In 2008 \cite{kaur08} claimed that  the QPO
frequency evolution during the previous 22 years changed from a positive to a negative
trend, somewhat coincident with the June 1990 torque reversal in this
source. \cite{orlandini98}  inferred a neutron star magnetic field in the range
(2.4--6.3)$\times$10$^{12}$ G, based on the idea that the
quasiperiodic oscillation frequency is due to the beating
between the pulse frequency  and the Keplerian motion at the
magnetospheric radius. To compute this magnetic field
range a source distance of 5--13 kpc was assumed. This distance
range was obtained from measurements of the optical and X--ray fluxes,
assuming that LMXB accretion disks have a very effective X--ray albedo
\citep[and references therein]{Chakrabarty98}.  Using a $\sim$37 keV absorption cyclotron feature found in the 0.1--200 keV  {\it BeppoSAX} spectrum, \cite{orlandini98}  obtained a value for
the magnetic field of $3.2 (1+ {\it z}) \times 10^{12}$\,G, where {\it z} is the
gravitational redshift. The X--ray  broad-band continuum was fitted
with a low--energy absorption, a blackbody, a power law and a high energy cutoff.

\begin{figure}[!th]

\includegraphics[width=8cm,height=7cm]{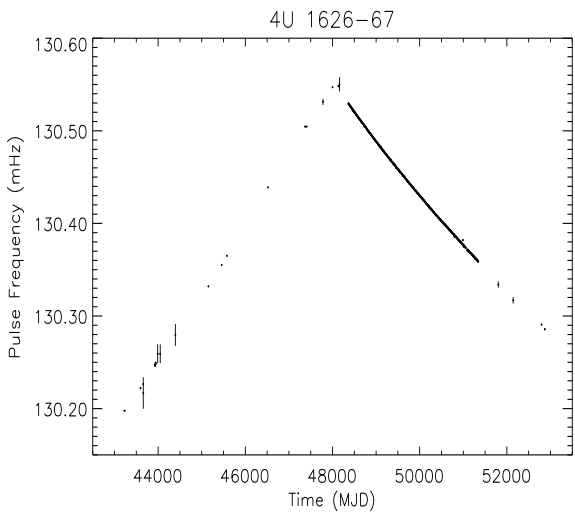}
 \caption{Pulse frequency history of 4U\,1626--67, showing  all available historical
 data from 1997 to 2003. The 1990 June ($\sim$MJD 48000) torque reversal is clearly seen.ff
\label{freqhist}}
\end{figure}


For more than a decade after the discovery of pulsations \citep{Rappaport77} the source underwent steady spin--up at a mean rate of $\sim \dot{\nu}=8.5 \times 10^{-13}$\,Hz\,s$^{-1}$ \citep{chakrabarty97} (see Fig~\ref{freqhist}). Monitoring of the source by the Burst and Transient Source Experiment (BATSE) on board the Compton Gamma Ray Observatory ({\it CGRO}) starting in  April 1991,  found the pulsar spinning down, implying a changed sign in the accretion torque \citep{Wilson93,Bildsten94}. During the 7 years after the first torque reversal, the pulsar  spun--down at a rate of $\sim \dot{\nu}=-7.2 \times 10^{-13}$\, H\, s$^{-1}$ \citep{chakrabarty97}.

Torque reversals have also been observed in the accreting pulsar systems Her X-1, Cen X-3, GX 1+4, OAO 1657-415, Vela X-1 \citep{bildsten97} and 4U\,1907+09 \citep{inam09}. The systems most similar to
4U 1626-67 are Her X-1 and Cen X-3, where a persistent accretion disk is know to
be present. Her X-1 like 4U\,1626-67 has a Roche-lobe filling low mass companion, where matter
from the companion flows from the L1 point into an accretion disk \citep{reynolds97}. Cen X-3 has a high mass companion which almost overflows its Roche-lobe with a focused wind from
companion flowing into the accretion disk \citep{day93}. Studies of accretion torque in Her X-1 are made difficult by the obstruction caused by the warped accretion disk which is
viewed from near edge-on, resulting in attenuation or complete eclipsing of the pulsar \citep{petterson91}, make flux measurements unreliable probes of the mass accretion rate. In Cen X-3 the
dense wind results in a large column density, with source occasionally being
completely obscured.  In GX\,1+4 \citep{hinkle06}, OAO\,1657-415 \citep{chakrabarty02}, 4U\,1907+09 \citep{fritz06} and Vela X-1 \citep{dupree80} the optical
companion underfills its Roche lobe and accretion proceeds from a wind,
with perhaps the transient formation of an accretion disk if the accreting
angular momentum is large enough. Vela X-1 is the prototype for a super-giant
wind fed system, where accretion proceeds by direct capture from the wind. In
this case the transfer of angular momentum is very inefficient. Detailed timing
analysis show that the frequency history of Vela X-1 is consistent with a random walk, or
equivalently the power-spectrum of the torque is white noise \citep{deeter89}. However, rare short high-flux states have been observed where transient disk accretion may occur \citep{krivonos03}. It is likely transient accretion disks occur in GX\,1+4, OAO\,1657-415, 4U\,1907+09 and play
an essential part there frequency histories. Understanding the torque behavior of
these systems will require characterization of the companions wind, and detecting the
pretense or absence of an accretion disk in addition to understanding the flow
of angular momentum between that disk and the neutron star. 4U\,1626-67 in contrast
provides a cleaner laboratory for investigating the flow of angular momentum.

We present a long term timing analysis using all the available \fermigbm data since
its launch in 2008 June 11 and over 5 yr of hard X-ray \swiftbat data from
2004 up to 2009. Spectral analysis was also carried out for 4U\,1626--67, in
order to see  changes   during  this second detected torque reversal   that could
help us to better  understand the physical mechanisms involved in this process.

\section{\fermigbm}

\subsection{OBSERVATIONS}

Since  2008 June 11  \f   has been continuously  monitored by the
{\it Gamma-ray Burst Monitor} (GBM)\citep[submitted]{gbmpaper}, on board the
{\it Fermi} observatory. Timing analysis was carried out with GBM  CTIME data,
with  8 channel spectra every 0.256 seconds. The total exposure time was
$\sim$13.75 Ms.

The  GBM is an all-sky instrument sensitive to X--rays and gamma rays with energies
between $\sim$8 keV and $\sim$40 MeV.  GBM includes 12 Sodium Iodide (NaI)
scintillation detectors and 2 Bismuth Germanate (BGO) scintillation
detectors. The NaI detectors cover the lower part of the energy range,
from 8  keV to about 1 MeV. The BGO detectors cover the energy range of
$\sim$150 keV to $\sim$40 MeV. Only data from the
NaI detectors were used in the analysis presented in this paper.

\subsection{TIMING ANALYSIS AND RESULTS}


\begin{figure}[!th]
\includegraphics[width=7.8cm,height=5.8cm]{Fig2a.eps}\\
\includegraphics[width=8.15cm,height=6cm]{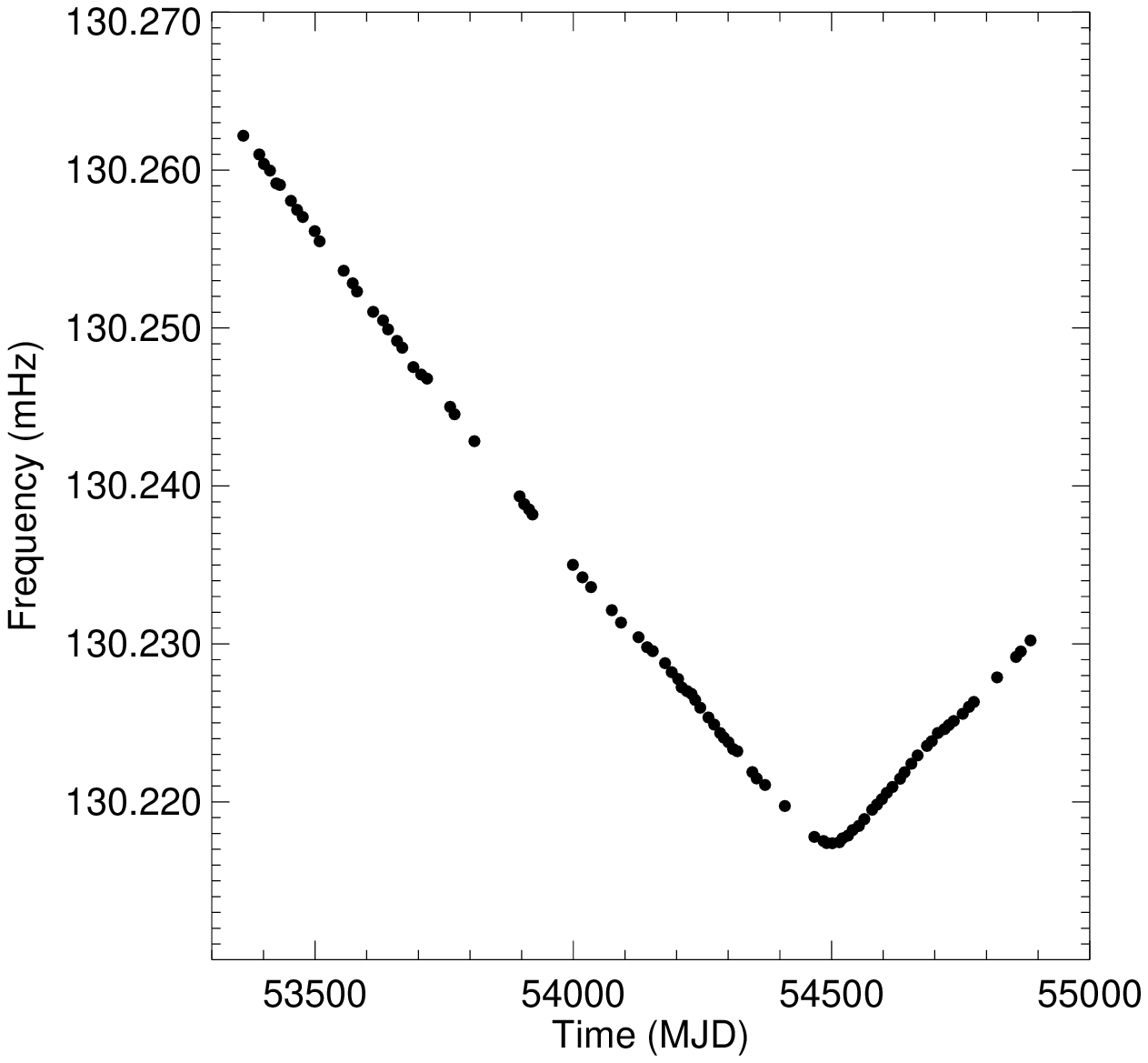}
 \caption{Top panel. \fermigbm pulse frequency measurements of 4U\,1626--67 since 2008 August.
 A change in the sign of the torque was found after 18 years of the source spinning down.
 Bottom panel. \swiftbat pulse frequency history  covering this second reversal
(from 2004 Oct to the present time). Error bars are smaller than the plotted symbols.
\label{pollas1}}
\vspace{0.2cm}
\end{figure}


\begin{figure}[!th]
 \includegraphics[width=7.8cm,height=12cm]{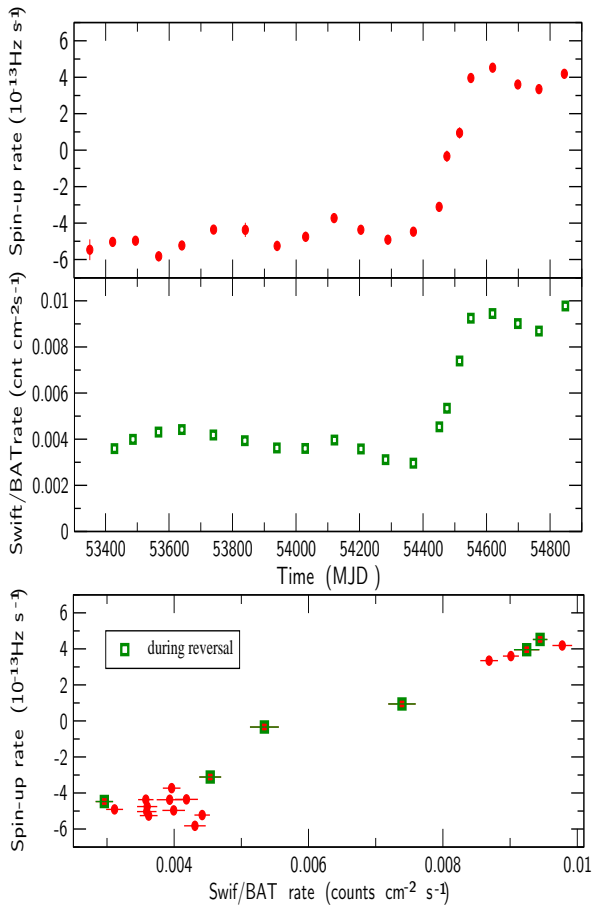}
 \caption{Top panel. \swiftbat spin--up rate history of \f. Middle panel. Average 15--50 keV BAT count rate vs. time. Error bars are  smaller than the plotted symbols. Bottom panel. BAT count rate vs. spin--up rate for all the period (circles). A correlation pattern is observed specially during the torque reversal (only square symbols).  \label{fdot}}
\vspace{0.3cm}
\end{figure}



The analysis of the Fermi/GBM data is complicated by Fermi's
continuously changing orientation. All intervals of CTIME data from the 12 NaI detectors
are selected for analysis where the high voltage is on, excluding those containing
high voltage transients, phosphorescence events, rapid spacecraft slews,
South Atlantic Anomaly induced transients, electron precipitation events
and gamma-ray bursts. Source pulses are then separated from
the background by fitting the rates in all detectors with a
background model, and subtracting the best fit model. This model includes bright sources
and their changing detector responses, including Earth occultation steps, and
quadratic spline functions which account for the remaining long-term background trends.  The
spline models have statistical constraints on the changes in second derivative
between spline segments to control the model stiffness. These fits are made jointly across
detectors (with common bright source fluxes) but separately for each channel of
the CTIME data. Then we combine the residuals over detectors
with time dependent weights which are proportional
to the predicted (phase averaged) count rates from the pulsar.
Short intervals ($\sim$300s) of these combined residuals
are then fit with a constant plus a Fourier expansion to determine a
pulse profile. The profiles are divided into six day intervals and the pulse frequency and mean
profile determined in each interval with a search of pulse frequency for the maximum of the Y$_{n}$ (n=2)  statistic \citep{finger99}. Y$_{n}$ was formulated for finding a pulse frequency from a series of pulse profiles, each represented by a finite Fourier expansion, and accounts for possible frequency dependent non-Poisson noise.



Our  monitoring of \f  with {\it Fermi}/GBM starting in 2008 August,  found the pulsar
again spinning--up rather than spinning--down. Fig~\ref{pollas1}
shows the pulse frequency history using
data from this monitoring. \f  seems to be increasing in $\dot{\nu}$. Follow-up {\it Fermi}/GBM observations  confirm that the pulsar it  is currently spinning--up  at a mean rate of $\sim\dot{\nu}$=4$\times10^{-13}$ Hz s$^{-1}$.

\section{\swiftbat}

\begin{table*}
\begin{scriptsize}
  \caption{ {\it RXTE}/PCA and HEXTE CONTINUUM SPECTRAL FITS$^{mod}$ }
  \begin{tabular}{lllllllllll}
  \hline\noalign{\smallskip}

  \hline\noalign{\smallskip}
  Observation   & $\alpha^*$  & E$_{cut}^{\dag}$       & E$_{Fold}^{\dag}$           & Gaussian$^{\dag}$   & Gauss.
                     & Gauss.         &T$_{BBody}^{\dag}$    &  norm  & Flux$^{***}$         &$\chi_r^{2}$(DOF)  \\
       (MJD)     &         &    &    &     &  $\sigma^{\dag}$
                  & norm$^{**}$   & T$_{Bremss}^{\dag}$        &   $^{BBod^1}_{Brem^2}$       &            &  \\
  \hline\noalign{\smallskip} \hline\noalign{\smallskip}

93431--01--01--00   & 0.75(2)& $18.2{+0.1\atop -0.3}$   & 8.5(4)  & $6.05{+0.5\atop -0.17}$& $1.6{+0.3\atop -0.2}$ & $2.4{+4\atop -0.8}$     &$0.615{+0.006\atop -0.018}$ &    0.0013      & 1.01(8) &1.15(114)      \\

    (54530)      &$0.74(2)$&$18.2{+0.3\atop -0.2}$ &$8.5{+0.3\atop -0.4}$& $6.42{+0.17\atop -0.4}$ &$1.4{+0.1\atop -0.2}$ &$1.5{+1.2\atop -0.6}$ & $1.74{+0.12\atop -0.14}$ & 0.108   & 1.01(1) & 1.15(114)   \\

  \hline\noalign{\smallskip}

 93431--01--02--00  & $0.71{+0.06\atop -0.04}$ & $17.90{+0.19\atop -0.3}$  & $8.4{+0.5\atop -0.6}$    & $6.2{+0.6\atop -0.4}$ & $1.5{+0.4\atop -0.2}$ & $2.23{+1.9\atop -0.08}$  & $0.654{+0.04\atop -0.009}$      &   0.0013      &1.006(12)& 1.29(114)\\

   (54538)      & $0.71{+0.04\atop -0.05}$  & $17.96{+0.14\atop -0.2}$  &8.4(6)  &6.9(2) &$0.5{+0.4\atop -0.3}$  & $0.396{+0.14\atop -0.015}$ &$2.47{+0.3\atop -0.08}$  &  0.068    &1.004(9)  &    1.28(114)\\

   \hline \noalign{\smallskip}

   \end{tabular}
    $^{mod}$WABS (GAUSSIAN+ (BLACKBODY or  BREMSSTRAHLUNG) +POW) HIGHECUT ($N_{\rm H} = 1.3\times10^{21}$\,cm$^{-2}$\,fixed; Uncertainties: 3$\sigma$\,level.)\\
    $^*$ Photon Index \\
    $^{\dag}$  keV\\
    $^{**}$    $\times 10^{-3}$ \\
    $^1$  (Lumin/$10^{39}$\,erg\,s$^{-1}$)(d/$10kpc)^{-2}$\\ 
    $^2$  $3.05\times 10^{-15} (4 \pi d^2)^{-1} \times$Emission measure   \\
    $^{***}$ $\times 10^{-9}$\,erg\,cm$^{-2}$\,s$^{-1}$    (2--100 keV)\\
\label{specfits}
\end{scriptsize}
\end{table*}


\subsection{OBSERVATIONS}

 The Swift Gamma-ray mission \citep{Gehrels04} was launched  on
2004 November 20. The hard X--ray (15--150 keV) Burst Alert Telescope
(BAT) on board {\it Swift} monitors the entire sky searching  mainly for
GRBs.   While searching for  bursts, BAT also accumulates a hard X--ray survey of the
entire sky covering 2 sr at any particular time. Therefore it produces continuous streams of {\it rate} data. The  {\it rate} data include {\it quadrant rates} (1.6 sec sampling; four energy bands; four separate spatial quadrants),  not background
subtracted.  For the present study we  have analyzed  more than 4  years
of BAT {\it quadrant rates} observations  when  \f \,was visible
(total exposure time $\sim13$\,Ms). For the hardness ratio analysis
we used count rates from the \swiftbat transient monitor results
provided by the \swiftbat  team\footnote{http://heasarc.gsfc.nasa.gov/docs/swift/results/transients}.

\subsection{ TIMING ANALYSIS  AND  RESULTS}

A similar procedure was followed for the \swiftbat {\it quadrant rates}
timing analysis. Initial good time interval (GTI) files are obtained using the {\it maketime}
ftool (heasoft-6.6.1)\footnote{http://heasarc.gsfc.nasa.gov/docs/software.html}. Then a filtered version  of the {\it quadrant rates}
is obtained, rejecting those  times when the source  is
below the horizon, to then finally be  barycentered using the ftool {\it barycorr}.
Data are then inspected and cleaned as in the previous section.   With the
ftool  {\it batmasktgimg} the pixel exposure fraction for each quadrant is
computed for the center of  each (refined)   GTI interval.  Pulse
profiles for  each good  GTI interval are computed. First the rates
for each quadrant  are fit to a  quadratic+Fourier expansion. Then
the Fourier coefficients are combined using the quadrant  exposures
 to produce mean profiles (with units of  counts s$^{-1}$ cm$^{-2}$). In a
final stage,  the  Y$_{n}$ (n=2) statistic is again used in intervals of 35
days and a frequency search for pulsations is carried out.  The spin rates were
computed by  fitting a linear function to the frequencies, which were divided into 21 time intervals.


\swiftbat observations allowed us to cover  the evolution of this
second torque reversal. We found that the pulsar
spun--down  at a mean rate of
$\sim \dot{\nu}=-4.8 \times 10^{-13}$\,Hz\,s$^{-1}$ until the source reversed
torque. Fig~\ref{pollas1}  shows that the transition took place at
around MJD 54500  (2008 Feb 04) and lasted  approximately  150 days.   In the bottom
panel of Fig~\ref{fdot} we can see that  there is a strong correlation between the
\swiftbat count rate and the spin--up rate especially during the reversal. It is not
known if this occurred  in the  1990  torque reversal because of the scarce
number of observations. In order to see any other possible change, we created
pulse profiles in the 15--50 keV  band (two harmonics were selected). We have not
observed any significant change in pulse shape, not even
during the reversal. They all are  single--peaked and sometimes not
entirely symmetric as was recently reported by \cite{krauss07}.

\section{{\it RXTE}}

\subsection{OBSERVATIONS}

The Rossi X--ray Timing Explorer ({\it RXTE}) \citep{bradt93} carries 3 instruments on board. The Proportional
Counter Array (PCA) \citep{Jahoda96} is sensitive from 2--60 keV. The
High Energy X-ray Timing Experiment (HEXTE) \citep{Gruber96} extends
the X-ray sensitivity up to 200 keV. Monitoring the long-term behavior
of some of the brightest X--ray sources, the All Sky Monitor
(ASM) \citep{levine96} scans most of the sky every 1.5 hours at
2--10 keV. Two {\it RXTE}/PCA observations from 2008 March 5 and 13
were used (ID 93431--01--01--00 and 93431--01--02--00; 7.174 ksec).  For spectral analysis
we selected PCA Standard--2 data which contains 129--channel
range spectra taken every 16 seconds and  HEXTE Standard Modes (Archive) data  which contains Spectral Bin (64-bin spectra produced every 16s). For the long--term hardness  ratio analysis we used  the ASM daily flux averages in the 1.5--12 keV energy range from the HEASARC  archive\footnote{http://heasarc.gsfc.nasa.gov/docs/archive.html}.

\subsection{SPECTRAL ANALYSIS AND RESULTS}

{\it RXTE}/PCA (2.5--20 keV) and  HEXTE (18--100 keV) spectra were fitted in XSPEC 11.3.2  with two models   used by \cite{pravdo79}.  
Using these models  allows us to compare our spectral study  
with previous works  by \citep{pravdo79,orlandini98,krauss07,jain09} and update the  long-term  X--ray flux history of 
4U\,1626--67  relative to the flux measured by {\it HEAO 1} (Chakrabarty et al. (1997); Krauss et al. (2007))  .
The first  model  includes a low-energy absorption, a blackbody component, a power law and a high-energy  cutoff at 
$\sim$20 keV  (WABS (GAUSS+BBODY+POWLAW) HIGHECUT).  A broad line near 6.5 keV  significantly improves the present 
fit and  indicates the presence of an iron line, also suggested by \cite{pravdo79} in their ( 0.7--100 keV) spectral 
analysis of this source. The column density of cool material in the line of sight was fixed in our study since it could 
not be constrained. A value of $1.3 \times 10^{21}$\,cm$^{-2}$ was selected from \cite{krauss07}.
The spectral parameters obtained are shown in Table 1.  Reprocessing of photons either at the base of the accretion column 
or the inner edge of the accretion disk  might  explain  the

\begin{figure}[!h]
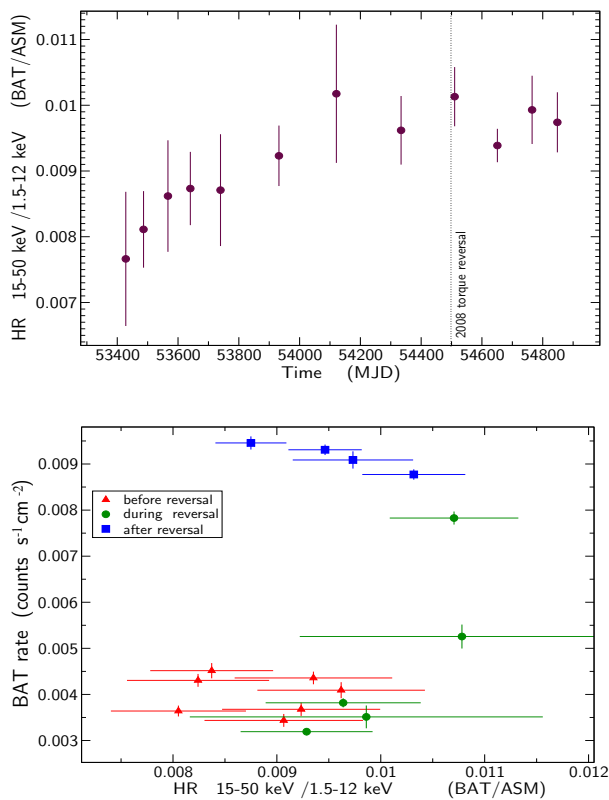

\vspace{0.1cm}
 \includegraphics[width=7.89cm,height=5cm]{Fig4a.eps}\vspace{+0.5cm}\\
 \includegraphics[width=8cm,height=5cm]{Fig4b.eps}

\caption{Top panel. Long term hardness ratio analysis of 4U\,1626--67. \swiftbat count rates (15--50 keV) were selected as the hard band  and {\it RXTE}/ASM count rate (1.5--12 keV) as the soft band. Bottom panel. Hardness--intensity diagram. During the reversal a transition from hard to soft is seen. \label{hr}}
\vspace{0.4cm}
\end{figure}


origin on the blackbody component. From the blackbody flux and temperature we obtain an emission area of 
$9\times 10^{12}$\,cm$^{-2}$  (assuming a source distance of 10 kpc), which is inconsistent with either of those possibilities.    
We fit in addition the same model with a bremsstrahlung instead of a blackbody component , obtaining  a 
compatible fit. Table~\ref{specfits}  summarizes the  spectral  
parameters obtained. Fluxes for the first model in the 2--10 keV,  2--20 keV and 2-50 keV bands are 3.24(2)$\times$10$^{-10}$, 
7.003(11)$\times$10$^{-10}$  and 9.98(2)$\times$10$^{-10}$ erg cm$^{-2}$s$^{-1}$, for the first {\it RXTE} observation 
(similar values for the second one).

\section{HARDNESS RATIO ANALYSIS}

Fig~\ref{hr} (top panel) shows the hardness ratio (HR) analysis carried
out for this source. The HR was defined  as the ratio 15--50keV/1.5--12keV
(BAT/ASM). To reduce large uncertainties the light curves were rebinned and
 then the HR were computed.  Like all hardness ratios, these are instrumentally
 dependent.  We can see that there is a smooth hardening
evolution of the source before the reversal, although  from this figure
we cannot observe any dramatic change a posteriori.

In the bottom panel of  Fig~\ref{hr} we have replaced the intensity with the BAT count rate
in order to perform Hardness--intensity diagram (HID). This allow us to study  the  long-term 
spectral variability of  \f,   including  the transition,  since the 2 RXTE  observations do not 
provide us  any direct comparison between before and after the torque reversal. From that figure 
we can see that there is a transition from
hard  to soft during this new reversal of 4U\,1626-67.

\section{DISCUSSION}

\begin{figure}[!h]
\vspace{0.5cm}
\includegraphics[width=8cm,height=7.1cm]{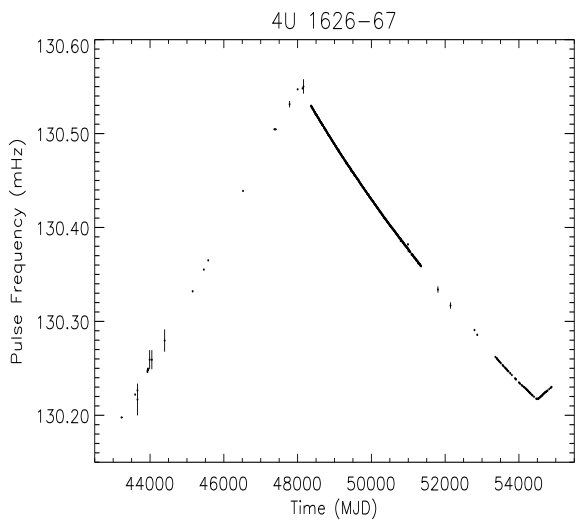}
\caption{Pulse frequency history of 4U\,1626--67 from 1997 up to 2009. The 1990 June and the 2008 Feb reversals are clearly seen. \label{allfreqhist}}
\end{figure}

The discovery of the torque reversal in 2008 (see Fig~\ref{allfreqhist}) has shown the
spin behavior of \f to be more complicated than previously thought. While the continuous
decrease of total X-ray flux during almost two decades was expected to bring the source
into quiescence \citep{krauss07}, a new spin reversal with a rapid increase of the flux
occurred. This inconsistency reopens discussion on physical processes governing the spin
evolution of the pulsar, and  raises a question about the nature of the new torque
reversal.

All previous studies of \f were focused on modeling the spin-up torque
applied to the neutron star from the accreted material. It was widely believed that the
spin behavior of the pulsar depended mainly on variations of the mass accretion rate onto
the stellar surface and therefore, the rate of mass transfer between the system
components. An attempt to explain the torque reversal in 1990 (the spin-up to spin-down
transition) in terms of variable equilibrium period \citep{ghoshlamb79} has been
made by \cite{vaughan97}. The neutron star phase transition in their model is associated
with a substantial decrease of mass accretion rate and, possibly, change of the

structure of the accretion flow beyond the magnetosphere. This scenario is based on some
evidence for a connection between torque and X-ray luminosity in \f and a hardening of
the energy spectrum after the torque reversal. They have emphasized, however, that this
approach encounters major difficulties explaining the observed behavior of QPOs, which
have been interpreted in terms of the magnetospheric beat-frequency oscillations  \citep{QPO}. Their
analysis suggests that either this interpretation of QPOs is incorrect or the torque
reversal in 1990 occurred as the magnetospheric radius of the neutron star, $r_{\rm m}$,
had increased beyond its corotation radius, which for the parameters of \f is
 \citep{lambpethickpines73}

 \be
r_{\rm c} = \left(\frac{GM_{\rm ns}}{4 \pi^2 \nu^2}\right)^{1/3}  \simeq 6.7 \times
10^8\ m^{1/3} P_{7.7}^{2/3}\ {\rm cm}.
 \ee
Here $m$ and $P_{7.7}$ are the mass and spin period of the neutron star in units of
$1.4\,M_{\sun}$ and 7.7$\,s$.  A dramatic change in the power spectra between the last observation of \f during the spin-down phase (2003)  \citep{kaur08}, and observations made soon after the new torque reversal has been recently reported by \cite{jain09}, with the 35--48 mHz QPO no longer being present, and wide shoulders on the pulse fundamental appearing. They claimed that the observed behavior of the source cannot be a simple case of increased mass transfer rate, but is also a change in the accretion flow parameters.

Analyzing the evolution of the source energy spectrum and possible correlation between
the torque and X-ray luminosity of the pulsar,  \cite{yi99} proposed a scenario in which
the torque reversal in 1990 is associated with a state transition of the accretion disk to a
geometrically thick, hot and, possibly, sub-Keplerian phase. Following this idea one
could associate the 2008 torque reversal with an inverse transition of the disk into its
previous geometrically thin Keplerian phase. However, the reason for such a transition
is rather unclear since the level of X-ray flux measured before and even after the 2008
reversal is smaller than that measured during the reversal in 1990. Furthermore, both
reversals have occurred at almost the same timescale (about 150\,days), which
significantly exceeds the dynamical timescale in the hot disk in which its transition to
the ground state is expected. A difficulty to fit in the transition timescales observed
in \f has also been mentioned by \citet{WijersPringle99}, who discussed a
possibility to explain the torque reversals in terms of the warped disk transition into
a retrograde regime.

\begin{figure}[!th]
\vspace{0.5cm}
\includegraphics[width=8cm,height=6.9cm]{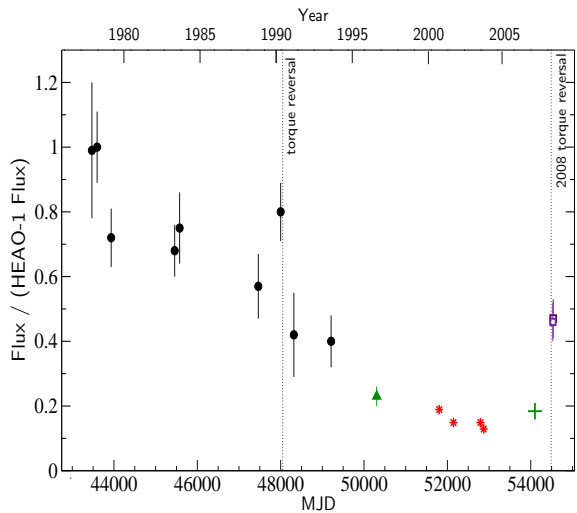}
 \caption{The X--ray flux history of 4U\,1626--67  relative to the flux measured by {\it HEAO 1},
in the same energy band, from previous works (Chakrabarty et al. (1997): circles; Orlandini et al.  (1998): triangle; Krauss et al. (2007): stars) and two recent {\it RXTE}/PCA observations (unfilled squares) in the 2--20 keV band. The cross point is inferred from PCA flux and the fractional change in the \swiftbat rate, since  no spectral changes during the transition have been observed in this work.\label{fluxhist}}

\end{figure}

A correlation between the torque applied to the neutron star in \f and X-ray flux of the
system in the above mentioned models has been adopted as one of the basic assumptions.
To test the validity of this assumption using data derived before 1993 was rather
complicated. This is illustrated in Figure~\ref{fluxhist}, which shows the \f X-ray flux
history. We can see all previous flux measurements and two {\it RXTE}/PCA recent values
from the present work (in the 2--20\,keV band). These values are relative to the flux
measured in 1978 by {\it HEAO 1} in the same energy band \citep[][and references
therein]{orlandini98, krauss07, chakrabarty97}. The cross point before the 2008 reversal
has been inferred by scaling the PCA fluxes according to the observed change (2.5
factor) in the \swiftbat rate, since  no spectral changes across the transition have
been observed according to the present work. It should be noted, however, that all these
relative flux values were computed in different energy bands, therefore the general
decreasing trend we see since 1977 might not be the real picture. Points could
misrepresent the HEAO\,1 0.7--60\,keV flux by a $\sim20\%$ due to spectral changes.
Moreover,  rather than a continuous decline (ignoring the 1990 point) \f might present a
flat behavior until the first reversal, then a sudden drop after that and a decreasing
tendency until the 2008 reversal, in which this source experienced a rapid increase of
flux. It therefore appears that the spin-down phase is the only suitable part of the
light curve for testing the correlation between the torque and mass accretion rate onto
the neutron star surface.

As seen from Figure~\ref{fluxhist}, the X-ray flux during the spin-down phase has
decreased by a factor of 2. This indicates that the mass accretion rate onto the surface
of the neutron star, $\dot{M}$, and, correspondingly, the spin-up torque applied to the
star \citep{pringlerees72},
 \be
K_{\rm su} = \dot{M} (GM_{\rm ns} r_{\rm m})^{1/2},
 \ee
during this phase have also decreased by at least the same value. If the spin-down
torque applied to the neutron star during this time were constant one would expect the
pulsar to brake harder at its fainter state close to the end of the spin-down phase.
However, observations show the situation to be just the opposite. The spin-down rate of
the neutron star during this phase has decreased from $|\dot{\nu}| \simeq 7 \times
10^{-13}\,{\rm Hz\,s^{-1}}$ \citep{chakrabarty97} to $5 \times 10^{-13}\,{\rm
Hz\,s^{-1}}$ (see Figure~\ref{pollas2}), implying that the pulsar was braking harder at
its brighter stage just after the torque reversal in 1990. According to the equation
governing spin evolution of an accreting neutron star,
 \be
2 \pi I \dot{\nu} = K_{\rm su} - K_{\rm sd},
 \ee
this means that the spin-down torque, $K_{\rm sd}$, during the spin-down phase has been
decreasing simultaneously with the spin-up torque but at a higher rate and, therefore,
the pulsar spin evolution during this time has been governed mainly by variations of
$K_{\rm sd}$ rather than $K_{\rm su}$ (here $I$ is the moment of inertia of the neutron
star). This conclusion seriously challenges the possibility of modeling the spin history
of \f solely in terms of variations of $\dot{M}$, and suggests that  the dramatic
increase of X-ray flux observed in 2008 torque reversal may be a consequence rather than
a reason for this event.

With the  lack of correlation between the X-ray flux and the torque applied to the
neutron star,  modeling of the spin-down torque appears to be the main target for
theoretical studies of the system. Unfortunately, this part of modeling of the
magneto-rotational evolution of neutron stars remains  so far a work in progress. The
canonical prescription for the spin-down torque \citep{Lipunov92},
 \be\label{ksd}
K_{\rm sd} = k_{\rm t} \frac{\mu^2}{r_{\rm c}^3},
  \ee
in our case turns out to be rather ineffective. The dipole magnetic moment of the
neutron star, $\mu$, during the spin-down phase obviously remains constant and the
corotation radius, changes only by 0.16$\%$ ($\Delta r_{\rm c} \la 1.1 \times 10^6$\,cm)
as it follows from the observed changes of the pulsar frequency ($|\Delta \nu| \simeq
3.3 \times 10^{-4}$).  Hence, any variation of $K_{\rm sd}$ under these conditions
proves to be determined by the dimensionless parameter $k_{\rm t} < 1$, which in the
case of disk accretion is just the $\alpha$ parameter at the inner radius of the disk
\citep[][and references therein]{Lipunov92}. However, it appears to be rather difficult
to combine the assumption about significant variation of turbulence at the inner radius
of the disk with the extremely low level of noise strength observed in \f
\citep{chakrabarty97}. Furthermore, different rates of variations of $K_{\rm sd}$ and
$K_{\rm su}$ during the spin-down phase indicates that either $k_{\rm t}$ is a
non-linear function of $\alpha$, or the expression~(\ref{ksd}) is oversimplified,  and
in the case of 4U\,1626--67 is ineffective.  As recently shown by \citet{Perna06}, the
spin-down torque proves to be a strongly dependent on the relative velocity between the
disk and magnetosphere and its value significantly varies even for a relatively small
variations of the mass-transfer rate if the interaction between that disk and the
magnetosphere occurs at the corotation radius of the star. They proposed a model where
simultaneous with accretion from a disk onto the neutron star some material from near
the disk -- magnetosphere boundary is ejected and either escapes from the system or is
recycled back into the accretion disk. This results in a hysteresis-type limit cycle
where slow changes in the accretion rate from the companion into the accretion disk can
result in the rapid change in torque and luminosity. Their model predicts, however, that
the luminosity after a spin-down to spin-up torque reversal would be higher than the
luminosity after a spin-up to spin-down torque reversal, which is the opposite of what
occurred for 4U~1626-67 for this new reversal. Moreover, for 4U~1626-67 they predicted
the full spin-down/spin-up cycle would take thousands of years, again inconsistent with
the present observations.  It should be noted, however, that the cycle could be
interrupted by a sudden increase of the mass transfer rate in the disk. The brightening
of the pulsar observed during the torque reversal in 2008 does not allow us to exclude
this possibility.

 The above mentioned problems may indicate that the torque applied to the
star should be treated in a different way \citep[see, e.g.][]{Rappaport04}, or the
interaction between the disk and magnetosphere in 4U~1626-67 occurs in a
particular  region, in which small perturbations of pulsar parameters may lead to dramatic
changes in the torque and mass accretion rate. As shown by \citet{Anzer80,
Anzer83}, the boundary between the disk and magnetosphere is subject to Kelvin-Helmholtz
instability in a region specified with the condition
 \be\label{delta}
|r_{\rm c} - r_{\rm m}| \la \delta_{\rm c},
 \ee
 while apart of this region the interchange-type instabilities of the
boundary are suppressed. Here $\delta_{\rm c}$ is the thickness of region where the
velocity difference between the disk and magnetosphere is smaller than the sound
velocity of plasma in the disk, i.e. $|V_{\rm k}(r_{\rm m} \pm \delta_{\rm c}) - V_{\rm k}(
r_{\rm c})| \la V_{\rm s}$. Here $V_{\rm k}(r)=(GM_{\rm ns}/r)^{1/2}$ is the
Keplerian velocity, and $V_{\rm s}$ is the sound speed at the inner radius of the disk.
The value of $\delta_{\rm c}$ is model dependent, but eventually is of the order of the
disk thickness \citep{Anzer83, Lovelace95}. For the conditions of interest
$\delta_{\rm c} \sim 3 \times 10^6$\,cm, which is comparable with the amplitude of
variation of the corotation radius during the spin-down phase 1990-2008 inferred from our
observations. This represents some grounds to assume that the spin and intensity evolution
of 4U~1626-67 can be governed by variation of relative position of $r_{\rm c}$
and $r_{\rm m}$ in the region specified by condition~(\ref{delta}). Since $\Delta r_{\rm
c} \simeq(1/2) \delta_{\rm c}$ one could expect that the spin-up/spin-down cycle of the
pulsar is determined mainly by variations of its corotation radius, while the
magnetospheric radius remains almost constant. The recurrent time of the cycle in this case
would be $\sim 15-25$\,year, which is close to the observed value. Further analysis of
the corresponding scenario is, however, beyond the scope of this paper. Here we
would like to note only that 4U~1626-67 represents an exceptional case as an
accretion-powered pulsar with an extremely small torque noise. This indicates that the accretion
picture in this system may differ significantly from that realized in other pulsars.

Finally, the spectral evolution of 4U 1626-67 during the torque reversal differs from
 that expected in models which suggest significant changes of the accretion flow
structure in spin-up/spin-down transitions  \citep[e.g.][]{yi99,
WijersPringle99}. As seen from Figure~\ref{hr}, the spectrum becomes the hardest during
the reversal and the value of the hardness ratio before and after these events does not
differ significantly. This indicates that the recent torque reversal can be associated
with changes of physical conditions in the inner part of the disk or/and in the
region of its interaction with the magnetosphere rather than a significant change of the
accretion flow geometry. The errors of the observations are, however, too large for a
justification of particular transition model. A more precise spectral measurements of
the pulsar during its next spin-up/spin-down transition is, therefore, strongly desired.
Assuming the recurrent time of the transition to be about 18\,years one can suggest to
pay more attention to the pulsar in 2025--2028. Since the typical duration of the
transition is about 150\,days,  a regular monitoring of the pulsar, frequently enough to provide a 
spin-up rate measurement every two months, would prevent us from missing its next torque reversal.

\section{CONCLUSIONS}

We report on a discovery of a new spin-down to spin-up torque reversal in 4U~1626-67. It
occurred after about 18\,years of the pulsar's steadily spinning down and was centered
on 2008 Feb 4. The transitions was lasted $\sim$150 days and accompanied by an increase
in the {\it Swift}/BAT count rate of a 2.5 factor ($\sim$150$\%$). The pulsar spectrum
was harder during the torque transition than before or after. A strong correlation
between torque and luminosity is inferred only during the transition. The spin-up and
spin-down rates before and after the transition were almost identical ($\sim
\mid\dot{\nu}\mid=5 \times10^{-13}$\,Hz\,s$^{-1}$). However, the pulsar was braking
harder at the beginning of the spin-down epoch in 1990 than at its end in 2008.
Furthermore, the spin-down rate during this epoch was decreasing simultaneously with the
decreasing of the source X-ray luminosity. Finally, the spin-down to spin-up torque
reversal in 2008 has occurred at lower luminosity as the spin-up to spin-down torque in
1990. These properties cannot be explained with existing models and  appear to be a clue 
for further progress in understanding the mechanism governing the torque reversals in the 
accretion -powered pulsars.

\acknowledgments {Acknowledgments. A.C.A. thanks for the support of this project to the
Spanish Ministerio de Ciencia e Innovaci\'on through the 2008 postdoctoral program
MICINN/Fulbright under grant 2008-0116. N.R.I. acknowledges supported from NASA
Postdoctoral Program at NASA Marshall Space Flight Center, administered by Oak
Ridge Associated Universities through a contract with NASA, and support from Russian
Foundation of Basic Research under the grant 07-02-00535a. M.H.F. acknowledges support
from NASA grant NNX08AG12G.  We also want to thank all the {\it Fermi}/GBM team for its help.}

\bibliographystyle{bibstyles/astron}
\bibliography{elif}

\end{document}